\documentclass[preprint,nofootinbib,floats,aps,superscriptaddress]{revtex4}
\usepackage{graphicx}

\begin{document}

\def\OMIT#1{{}}
\def\lqcd{\Lambda_{\rm QCD}}
\def\d{{\rm d}}
\def\FD{{\cal F}}
\def\FDs{\FD_*}
\def\FDt{\FD_{(*)}}

\preprint{\vbox{ \hbox{LBNL--49173} \hbox{UCSD/PTH 01--20} 
  \hbox{hep-ph/0111392} }}

\title{Heavy quark symmetry in $B\to D^{(*)}\ell\bar\nu$ spectra}

\vspace*{1cm}

\author{Benjam\'\i{}n Grinstein}

\affiliation{Physics Department, University of California at San Diego, 
  La Jolla, CA 92093}

\author{Zoltan Ligeti}
\affiliation{Ernest Orlando Lawrence Berkeley National Laboratory \\
  University of California, Berkeley, CA 94720 \\ $\phantom{}$}

\begin{abstract}

We calculate heavy quark symmetry breaking in the slopes and curvatures of the
$B\to D^{(*)}\ell\bar\nu$ spectra at zero recoil, including the order
$\alpha_s^2\beta_0$ corrections.  We point out that the theoretical
uncertainties in the differences between $B\to D$ and $B\to D^*$ slopes and
curvatures are smaller than in the deviations of the slopes and curvatures
themselves from their infinite mass limits.  We find that the central values of
the current experimental results for the difference of the slopes differ from
our calculations when QCD sum rules are used to estimate subleading Isgur-Wise
functions.  A better understanding of the shapes of the $B\to
D^{(*)}\ell\bar\nu$ spectra may also help to reduce the error of $|V_{cb}|$
extracted from the zero recoil limit of $B\to D^*\ell\bar\nu$. We argue that
heavy quark symmetry requires that the same fitting procedure be used in the
experimental determinations of the shape parameters and $|V_{cb}|$ from the
$B\to D\ell\bar\nu$ and $B\to D^{*}\ell\bar\nu$ spectra.

\end{abstract}

\maketitle

\section{Introduction}

The determination of $|V_{cb}|$ from exclusive $B\to D^{(*)}\ell\bar\nu$ decays
is based on the fact that heavy quark symmetry~\cite{HQS} relates the form
factors which occur in these decays to the Isgur-Wise function, whose value is
known at zero recoil in the infinite mass limit. The symmetry breaking
corrections can be organized in a simultaneous expansion in $\alpha_s$ and
$\lqcd/m_Q$ ($Q = c,b$).  The $B\to D^{(*)}\ell\bar\nu$ decay rates are given
by
\begin{eqnarray}\label{rates}
{\d\Gamma(B\to D^*\ell\bar\nu)\over \d w} &=& {G_F^2 m_B^5\over 48\pi^3}\, 
  r_*^3\, (1-r_*)^2\, \sqrt{w^2-1}\, (w+1)^2 \nonumber\\[4pt]
&& \times \left[ 1 + {4w\over 1+w} {1-2wr_*+r_*^2\over (1-r_*)^2} \right]
  |V_{cb}|^2\, \FDs{}^2(w) \,, \nonumber\\[4pt]
{\d\Gamma(B\to D \ell\bar\nu)\over \d w} &=& {G_F^2 m_B^5\over 48\pi^3}\, 
  r^3\, (1+r)^2\, (w^2-1)^{3/2}\, |V_{cb}|^2\, \FD^2(w) \,, 
\end{eqnarray}
where $w = v\cdot v' = (m_B^2 + m_{D^{(*)}}^2 - q^2) / (2m_B m_{D^{(*)}})$ and
$r_{(*)} = m_{D^{(*)}}/m_B$.  Both $\FD(w)$ and $\FDs(w)$ are equal to the
Isgur-Wise function in the $m_Q \to\infty$ limit, and in particular $\FD(1) =
\FDs(1) = 1$, allowing for a model independent determination of $|V_{cb}|$.

The main theoretical uncertainties in such a determination of $|V_{cb}|$ come
from the value of $\FDt(1)$, and from the shape of $\FDt(w)$ used to fit the
data.  Such a fit will continue to be important, since the number of $B\to
D^*\ell\bar\nu$ events needed to measure $|V_{cb}|\, \FDs(1)$ with a
statistical error of order $(\lqcd/m_Q)^2$ scales parametrically as
$(m_Q/\lqcd)^7$.  This can be seen from Eq.~(\ref{rates}), making no 
assumption on the shape of $\FDs(w)$, by considering a bin at zero recoil with
width of order of the desired accuracy, that is, of order $(\lqcd/m_Q)^2$. 
Similarly, when unquenched lattice QCD calculations of $\FDt(1)$ will be
available with error $a$, the number of events needed to measure $|V_{cb}|\,
\FDt(1)$ with a comparable statistical error, without assumptions about the
shapes of $\FDt(w)$, will scale as $a^{-7/2}$ in $B\to D^*\ell\bar\nu$ and as
$a^{-9/2}$ in $B\to D\ell\bar\nu$.  Reliable unquenched lattice QCD results for
$\FDt(1)$ are likely to be available before comparable results for the
functional form of $\FDt(w)$. Constraining the shapes of $\FDt(w)$ will remain
important in the foreseeable future.

The zero recoil limit of $\FDt(w)$, including symmetry breaking corrections,
can be written schematically as
\begin{eqnarray}\label{F1}
\FDs(1) &=& 1 + c_A(\alpha_s) + {0\over m_Q} 
  + {(\ldots)\over m_Q^2} + \ldots \,, \nonumber\\*
\FD(1) &=& 1 + c_V(\alpha_s) + {(\ldots)\over m_Q} 
  + {(\ldots)\over m_Q^2} + \ldots \,.
\end{eqnarray}
The perturbative corrections, $c_A = -0.04$ and $c_V = 0.02$, have been
computed to order $\alpha_s^2$~\cite{Czar}, and the unknown higher order
corrections should be below the 1\% level.  The order $\lqcd/m_Q$ correction
to $\FDs(1)$ vanishes due to Luke's theorem~\cite{Luke}.  The terms indicated
by $(\ldots)$ in Eqs.~(\ref{F1}) are only known using phenomenological models
or quenched lattice QCD at present.  This is why the determination of
$|V_{cb}|$ from $B\to D^* \ell\bar\nu$ is theoretically more reliable for now
than that from $B\to D\ell\bar\nu$, although both QCD sum rules~\cite{LNN}
and quenched lattice QCD~\cite{lattice} suggest that the order $\lqcd/m_Q$
correction to $\FD(1)$ is small.  Due to the extra $w^2-1$ helicity
suppression near zero recoil, $B\to D\ell\bar\nu$ is also more difficult
experimentally.

Some of the order $\lqcd/m_Q$ corrections which enter $\FDt(w)$ also influence
ratios of form factors measurable in $B\to D^*\ell\bar\nu$ decay.  The
exclusive semileptonic $B\to D^* \ell\bar\nu$ decay rate is parameterized by
four form factors,
\begin{eqnarray}\label{formf}
{\langle D^*(v',\epsilon)|\, \bar c\, \gamma^\mu\, b\, |B(v)\rangle 
  \over \sqrt{m_{D^*}\,m_B}}
&=& i\, h_V\, \varepsilon^{\mu\alpha\beta\gamma} 
  \epsilon^*_\alpha v'_\beta v_\gamma \,, \nonumber\\*
{\langle D^*(v',\epsilon)|\, \bar c\, \gamma^\mu\gamma_5\, b\,\, |B(v)\rangle
  \over \sqrt{m_{D^*}\,m_B}}
&=& h_{A_1} (w+1)\, \epsilon^{*\mu} 
  - (h_{A_2} v^\mu + h_{A_3} v'{}^\mu)\, (\epsilon^*\cdot v) \,.
\end{eqnarray}
One linear combination is not measurable when the lepton masses are 
neglected.  The form factor $h_{A_1}$ dominates the rate near zero recoil.  
It is conventional~\cite{physrep} to define two measurable ratios of the form
factors
\begin{equation}\label{R12def}
R_1(w) = {h_V(w) \over h_{A_1}(w)}\,, \qquad
R_2(w) = {h_{A_3}(w) + (m_{D^*}/m_B) h_{A_2}(w) \over h_{A_1}(w)}\,.
\end{equation}
In the infinite mass limit $R_1(w) = R_2(w) = 1$, and deviations from this
limit measure certain combinations of the subleading Isgur-Wise functions as it
will be discussed it later.

Since there are several form factors in $B\to D^*\ell\bar\nu$, it is customary
to fit the shape parameters of $h_{A_1}$, together with the form factor ratios
$R_{1,2}$.  For the shapes of $\FD(w)$, $\FDs(w)$, and $h_{A_1}(w)$,
analyticity imposes stringent constraints between the slopes and curvatures at
zero recoil~\cite{BGL}.  It is convenient to write
\begin{eqnarray}\label{Fw}
\FD(w) &=& \FD(1) \left[ 1 - \rho_{\FD}^2\, (w-1) 
 + c_{\FD}\, (w-1)^2 + \cdots \right] , \nonumber\\[4pt]
\FDs(w) &=& \FDs(1) \left[ 1 - \rho_{\FDs}^2\, (w-1) 
 + c_{\FDs}\, (w-1)^2 + \cdots \right] , \nonumber\\[4pt]
h_{A_1}(w) &=& h_{A_1}(1) \left[ 1 - \rho_{A_1}^2\, (w-1) 
  + c_{A_1}\, (w-1)^2 + \cdots \right] .
\end{eqnarray}
A lower index $X$ will denote any of $\FD$, $\FDs$ or $A_1$. In the $m_Q\to
\infty$ limit, heavy quark symmetry predicts $\rho_X = \rho_0$ and $c_X = c_0$,
the slope and curvature of the Isgur-Wise function, respectively.  These
equalities are violated by corrections at order $\alpha_s$ and $\lqcd/m_Q$.  In
a linear fit to the data the expansions in~(\ref{Fw}) are terminated at linear
order in $w-1$, while in a quadratic (or ``free curvature'') fit the expansions
are terminated at order $(w-1)^2$. Higher order terms are small but
non-negligible, since the fitted range extends to $w = 1.59\ (1.50)$ in $B\to
D^{(*)}$ decay.  Unitarity constraints give a strong correlation between
$\rho_{\FDt}^2$ and the coefficients of the higher order terms in
$w-1$~\cite{BGL,CLN}, and also constrain the effect of the higher than
quadratic terms in $w-1$~\cite{CLN}.  For these reasons, an unconstrained
quadratic fit to the data is not equivalent to a unitarity constrained fit.

\begin{figure*}
\centerline{\includegraphics*[height=0.5\textwidth]{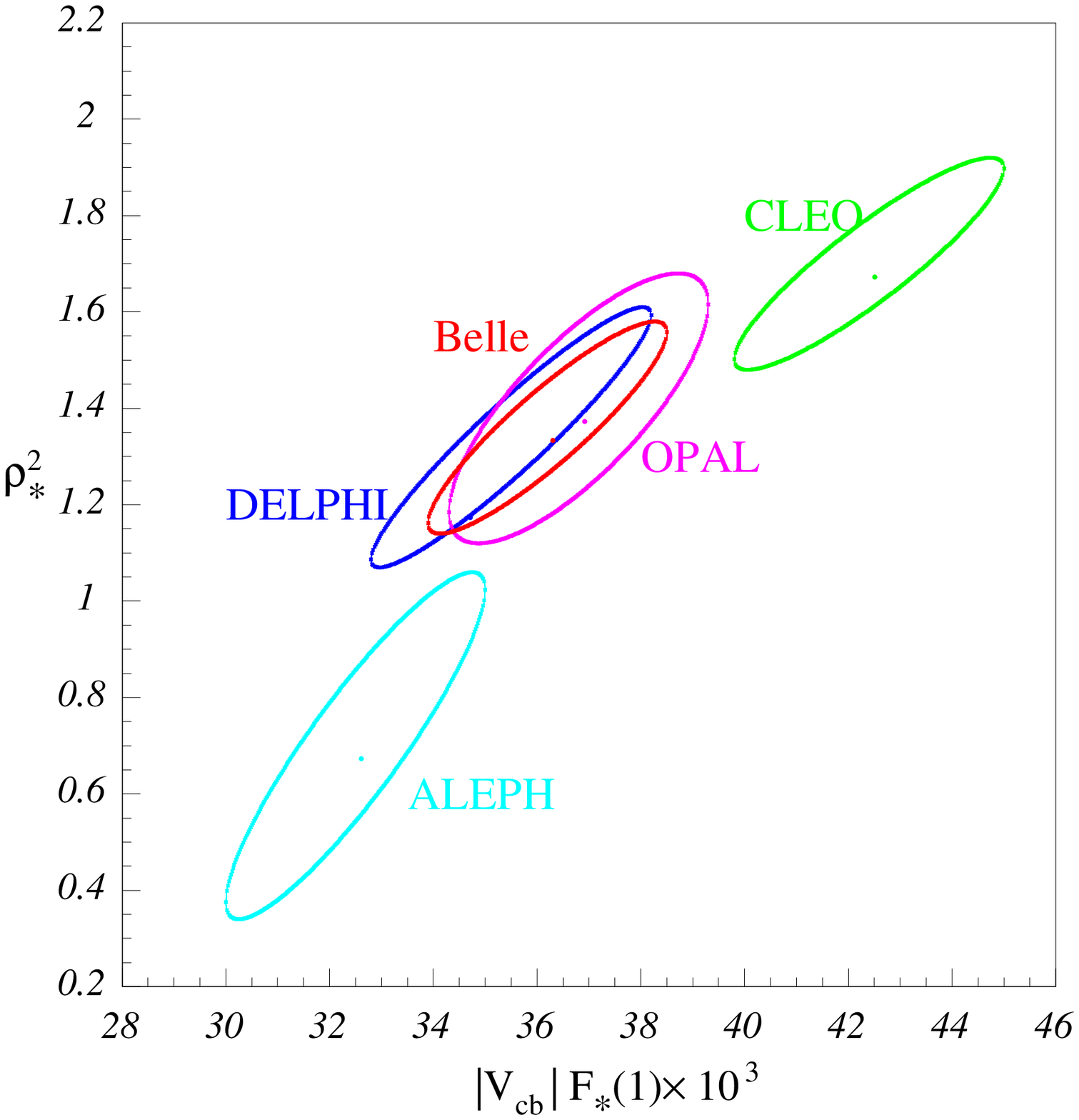} \hfill
  \includegraphics*[height=0.5\textwidth]{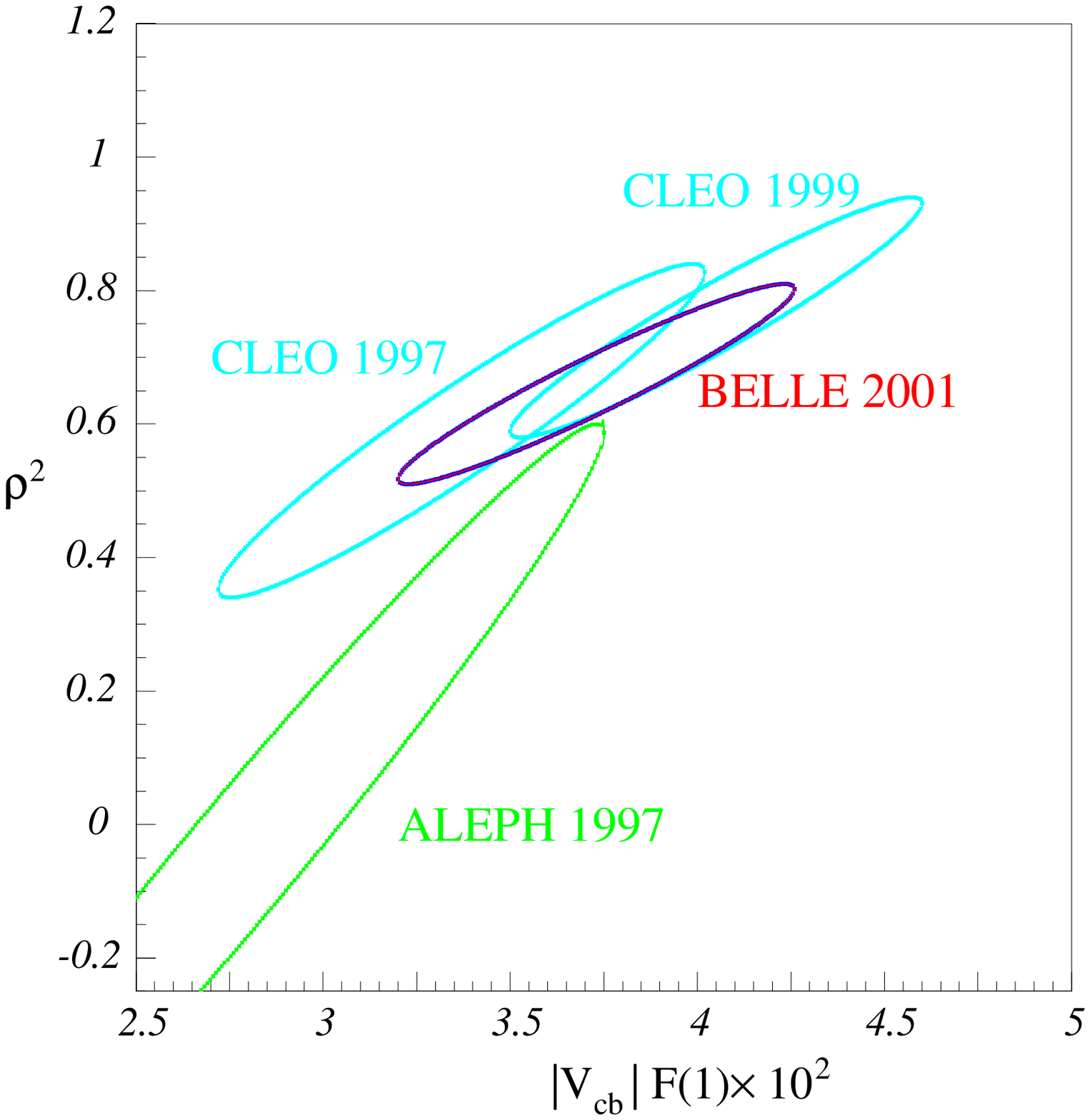}}
\caption{The measured values of $|V_{cb}|\, \FDs(1)$ and $\rho_{A_1}^2$
from unitarity constrained fits to $B\to D^*\ell\bar\nu$ decay (left), and
$|V_{cb}|\, \FD(1)$ and $\rho_{\FD}^2$ from linear fits to the $B\to
D\ell\bar\nu$ decay (right).  From Ref.~\cite{Belle}.}
\end{figure*}

\begin{table}[t]
\begin{tabular}{l||c|c} \hline\hline
\multicolumn{1}{c||}{Fitted slope parameter}  &  CLEO  &  BELLE \\ \hline\hline
~$B\to D^*\ell\bar\nu$, unitarity constrained fit to $\rho_{A_1}^2$~  &
  ~$1.67 \pm 0.11 \pm 0.22$ \cite{CleoDs}~  &
  ~$1.35 \pm 0.17 \pm 0.19$ \cite{BelleDs}~ \\
~$B\to D^*\ell\bar\nu$, linear fit to $\rho_{\FDs}^2$  &
  ~$0.98 \pm 0.09 \pm 0.07$ \cite{Cleoprivate}~  &
  ~$0.89 \pm 0.09 \pm 0.05$ \cite{Belleprivate}~ \\ \hline
~$B\to D\ell\bar\nu$, unitarity constrained fit to $\rho_{\FD}^2$  &
  ~$1.30 \pm 0.27 \pm 0.14$ \cite{CleoD}~  &
  ~$1.16 \pm 0.25 \pm 0.15$ \cite{BelleD}~ \\ 
~$B\to D\ell\bar\nu$, linear fit to $\rho_{\FD}^2$  &
  ~$0.76 \pm 0.16 \pm 0.08$ \cite{CleoD}~  &
  ~$0.69 \pm 0.14 \pm 0.09$ \cite{BelleD}~ \\ \hline\hline
\end{tabular} 
\caption{The most recent available CLEO and BELLE measurements of the slope parameters,
$\rho_X^2$, in $B\to D^{(*)}\ell\bar\nu$ decay.  The difference between
fitting to the unitarity constrained shapes of one or the other of
Refs.~\cite{BGL} or 
\cite{CLN} is very small.}
\end{table}

In the next section we compute the violations to the symmetry relations between
the slope and curvature parameters and find them to be small. Moreover, we
observe that in the present data the slope parameters $\rho_{\FDs}^2$ and
$\rho_{\FD}^2$ differ significantly if the spectra are fit linearly or
quadratically; see Table~I and Fig.~1. However, when present data is fit to the
functional form constrained by unitarity, the result is consistent with the
symmetry relation $\rho_{\FDs}^2 = \rho_{\FD}^2$.  Therefore, if in the future
the data supports the necessity of a unitarity constrained form for $\FDs(w)$,
then it will be equally necessary to use a unitarity constrained fit for
$\FD(w)$. 

Furthermore, to the extent that we can reliably estimate the deviations from
$\rho_{\FDs}^2 = \rho_{\FD}^2$, the accuracy in the extraction of $|V_{cb}|$
can be improved by a simultaneous fit to $B\to D^*$ and $B\to D$.  While the
value of $|V_{cb}|\, \FD(1)$ has larger uncertainty than $|V_{cb}|\, \FDs(1)$,
since $B\to D\ell\bar\nu$ is helicity suppressed near zero recoil, the error on
$\rho_{\FDt}^2$ is comparable as extracted from $B\to D^*$ and $B\to D$.  Since
there is a very strong correlation between $\rho_{\FDt}^2$ and $|V_{cb}|\, \FDt
(1)$, despite the fact that $\FD (1)$ is less well-known than $\FDs(1)$, a
better determined slope may have significant implications for the value of
$|V_{cb}|$ as it is extracted from the zero recoil limit of $B\to
D^*\ell\bar\nu$.  When precise unquenched lattice calculations of $\FDt(w)$
become available, consistency between the predictions and data for $\rho_X^2$
may be an important cross-check of the $|V_{cb}|$ determination.

\section{Analytic results}

We next give the theoretical predictions for the heavy quark symmetry violation
in $\rho_X^2$, $c_X$, and the form factor ratios $R_{1,2}$ at order
$\lqcd/m_Q$, $\alpha_s$ and $\alpha_s^2\beta_0$.  The first two corrections are
known in the literature, but the order $\alpha_s^2\beta_0$ terms, which
probably dominate the order $\alpha_s^2$ corrections (since $\beta_0 =
11-2n_f/3$ is large) are new.  These are required to predict heavy quark
symmetry breaking in the quantities under consideration at the $\sim 5\%$
level, and may become important in testing lattice results.  We can write the
slope parameters $\rho_X^2$ as
\begin{equation}\label{rho2rel}
\rho_X^2 = \rho_0^2 + \delta^{(\alpha_s)}_X
  + {\bar\Lambda\over2m_c}\, \delta^{(1/m)}_X \,,
\end{equation}
where $X=\FD,\, \FDs,\, A_1$ denotes the functions under consideration.  The
perturbative corrections can be calculated model independently and are
contained in $\delta^{(\alpha_s)}_X$, while $\delta^{(1/m)}_X$ contains the
order $\Lambda_{\rm QCD}/m_Q$ corrections to $\rho_X^2$.  The perturbative
corrections can be computed including the order $\alpha_s^2\beta_0$ corrections
from expanding Eqs.~(39) in Ref.~\cite{NS}.  We obtain
\begin{eqnarray}\label{deltaAs}
\delta^{(\alpha_s)}_{\FD} - \delta^{(\alpha_s)}_{\FDs} 
&=& {\bar\alpha_s\over\pi}\, \bigg[
  {8(1-3z+z^2)\over 9(1-z)^2} - {4z(1+z) \over 9(1-z)^3}\, \ln z \bigg]
  \nonumber\\*
&&{} + {\bar\alpha_s^2\over\pi^2}\, \beta_0\, \bigg[
  {8-17z+8z^2\over 27(1-z)^2} - {(1+z)(12-23z+12z^2) \over 54(1-z)^3}\, \ln z
  \bigg] \,, \nonumber \\
\delta^{(\alpha_s)}_{A_1} - \delta^{(\alpha_s)}_{\FDs} 
&=& {\bar\alpha_s\over\pi}\, \bigg[
  {4(1-z+z^2)\over 9(1-z)^2} + {2z(1+z) \over 9(1-z)^3}\, \ln z \bigg]
  \nonumber\\*
&&{} + {\bar\alpha_s^2\over\pi^2}\, \beta_0\, \bigg[
  {4+5z+4z^2\over 54(1-z)^2} - {(1+z)(12-37z+12z^2) \over 108(1-z)^3}\, \ln z
  \bigg] \,,
\end{eqnarray}
where $z = m_c/m_b$, and $\bar\alpha_s$ denotes the strong coupling
renormalized in the $\overline{\rm MS}$ scheme at the scale $\mu = \sqrt{m_b
m_c}$.  The terms of order $\alpha_s$ agree with Ref.~\cite{BLRW}.  Using $z =
0.29$, $\bar\alpha_s = 0.26$, and $\beta_0 = 25/3$, we obtain
\begin{equation}\label{slopenum}
\delta^{(\alpha_s)}_{\FD} - \delta^{(\alpha_s)}_{\FDs} = 0.079 + 0.046\,,
  \qquad
\delta^{(\alpha_s)}_{A_1} - \delta^{(\alpha_s)}_{\FDs} = 0.034 + 0.018\,,
\end{equation}
where the first and second terms come from the order $\bar\alpha_s$ and
$\bar\alpha_s^2 \beta_0$ corrections, respectively.  The apparent bad
convergence of the perturbation series in Eq.~(\ref{slopenum}) may be due to
the fact that they contain so-called renormalon ambiguities, and are only
well-defined physical quantities when the nonperturbative corrections discussed
next [see Eq.~(\ref{delta1m})] are included.  This assertion is supported by
the fact that we will find significantly better behavior when these series are
expressed in terms of a short distance mass in Sec.~III.

The corrections in $\delta_X^{(1/m)}$ depend on the four subleading Isgur-Wise
functions that parameterize first order deviations from the infinite mass
limit.  Using the notation of~\cite{physrep}, one finds from Ref.~\cite{BLRW}
\begin{eqnarray}\label{delta1m}
\delta^{(1/m)}_{\FD}- \delta^{(1/m)}_{\FDs} 
&=& \frac{5(1+z)}6\, + \frac{16}3\,\chi_2(1) - 16\chi_3'(1)
  + {1-2z+5z^2\over3(1-z)}\,\eta(1) + {2(1-z)^2\over1+z}\, \eta'(1)\,,
  \nonumber\\
\delta^{(1/m)}_{A_1} - \delta^{(1/m)}_{\FDs} 
&=& \frac{1+z}3 + \frac43\,\chi_2(1) + {1+z+2z^2 \over 3(1-z)}\,\eta(1)\,,
\end{eqnarray}
where prime denotes ${\rm d}/{\rm d}w$.  $\eta(w) \equiv \xi_3(w)/\xi(w)$
parameterizes the ${\cal O}(\lqcd/m_{c,b})$ corrections to the $b\to c$
current, and $\chi_{2,3}(w)$ describe the matrix elements of the time ordered
product of the chromomagnetic operator, $(g_s/2)\, \bar h_v \sigma_{\mu\nu}
G^{\mu\nu} h_v$, with the leading order current.  An important point is that
the poorly known function $\chi_1(w)$ which parameterizes matrix elements
involving the time ordered product of the kinetic energy operator, $\bar h_v
(iD)^2 h_v$, with the leading order current drops out from the differences in
Eq.~(\ref{delta1m}).  In general, $\chi_1$ does not effect any quantity
determined by heavy quark symmetry at leading order, such as the slope and
curvature differences in Eqs.~(\ref{delta1m}) and (\ref{deltacm}), and
$R_{1,2}$ in Eqs.~(\ref{R1exp}) and (\ref{R2exp}).  The reason is that $\chi_1$
enters all form factors in the combination $\xi(w) + (\bar\Lambda/m_c + 
\bar\Lambda/m_b)\, \chi_1(w)$, where $\xi(w)$ is the Isgur-Wise function.

It is straightforward to compute the heavy quark symmetry breaking corrections 
between $c_{\FD}$, $c_{\FDs}$, and $c_{A_1}$.  Similar to Eq.~(\ref{rho2rel}),
we define
\begin{equation}\label{crel}
c_X = c_0 + \Delta^{(\alpha_s)}_X
  + {\bar\Lambda\over2m_c}\, \Delta^{(1/m)}_X \,. 
\end{equation}
For the differences of $\Delta^{(\alpha_s)}_X$ we obtain
\begin{eqnarray}\label{deltac}
\Delta^{(\alpha_s)}_{\FD} - \Delta^{(\alpha_s)}_{\FDs}
&=& {2\bar\alpha_s\over 135\pi}\, \bigg[
  {47-148z+282z^2-148z^3+47z^4 \over (1-z)^4} +
  {5z(1+z)(1+6z+z^2) \over (1-z)^5}\, \ln z \bigg] \nonumber\\
&&{} + {\bar\alpha_s^2\over 405 \pi^2}\,\beta_0\, \bigg[
  {509 - 1241z + 1084z^2 - 1241z^3 + 509z^4\over 5(1-z)^4} \nonumber\\
&&{}\qquad\qquad\quad
  - {(1+z)(282 - 1277z + 2142z^2 - 1277z^3 + 282z^4)\over 4(1-z)^5}\, 
  \ln z \bigg] \nonumber\\*
&&{} + \rho_0^2\, \Big[ \delta^{(\alpha_s)}_{\FD} 
  - \delta^{(\alpha_s)}_{\FDs} \Big] \,, \nonumber\\[6pt]
\Delta^{(\alpha_s)}_{A_1} - \Delta^{(\alpha_s)}_{\FDs}
&=& {2\bar\alpha_s\over 27\pi}\, \bigg[
  {7-23z+12z^2-23z^3+7z^4 \over (1-z)^4} +
  {z(1+z)(1-12z+z^2) \over (1-z)^5}\, \ln z \bigg] \nonumber\\*
&&{} + {\bar\alpha_s^2\over 324 \pi^2}\,\beta_0\, \bigg[
  {44 - 83z - 134z^2 - 83z^3 + 44 z^4 \over (1-z)^4} \nonumber\\*
&&{}\qquad\qquad\quad 
  - {(1+z)(42 - 187z + 396z^2 - 187z^3 + 42z^4)\over 4(1-z)^5}\, 
  \ln z \bigg] \nonumber\\*
&&{} + \rho_0^2\, \Big[ \delta^{(\alpha_s)}_{A_1} 
  - \delta^{(\alpha_s)}_{\FDs} \Big] \,.
\end{eqnarray}
The order $\alpha_s$ piece in the first of these equations agrees with 
Ref.~\cite{CN}.  To this order, $\rho_0^2$ can be taken as any of
$\rho_{\FD}^2$, $\rho_{\FDs}^2$, or $\rho_{A_1}^2$.  With the previously used
values of $z$ and $\bar\alpha_s$, we obtain
\begin{eqnarray}\label{curvnum}
\Delta_{\FD}^{(\alpha_s)} - \Delta_{\FDs}^{(\alpha_s)} 
  &=& (0.074 + 0.043) + \rho_0^2\, (0.079 + 0.046) \,, \nonumber\\*
\Delta_{A_1}^{(\alpha_s)} - \Delta_{\FDs}^{(\alpha_s)} 
  &=& (0.058 + 0.031) + \rho_0^2\, (0.034 + 0.018) \,,
\end{eqnarray}
where the first and second terms in each parenthesis come from the order
$\bar\alpha_s$ and $\bar\alpha_s^2\beta_0$ corrections, respectively.  The
convergence of these series is again quite poor.  The order $\lqcd/m_{c,b}$
heavy quark symmetry breaking in the curvature differences is given by
\begin{eqnarray}\label{deltacm}
\Delta^{(1/m)}_{\FD}- \Delta^{(1/m)}_{\FDs} 
&=& {(1+z)(25-42z+25z^2)\over 36(1-z)^2} 
  + 4\chi_2(1)\, {1 - 6z + z^2\over 9(1-z)^2}
  - \frac{16}3\, \chi_2'(1) + 8\chi_3''(1) \nonumber\\*
&+& \eta(1)\, {5 - 28z + 18z^2 - 52z^3 + 25z^4 \over 18(1-z)^3}
  - \eta'(1) {1 - 2z + 5z^2 \over 3(1-z)} 
  - \eta''(1)\, {(1-z)^2\over(1+z)} \nonumber\\*
&+& \rho_0^2\, \Big[ \delta^{(1/m)}_{\FD}- \delta^{(1/m)}_{\FDs} 
  - \frac{16}3\, \chi_2(1) + 16\, \chi_3'(1) \Big] \nonumber\\[6pt]
\Delta^{(1/m)}_{A_1} - \Delta^{(1/m)}_{\FDs} 
&=& {2(1+z)(2-3z+2z^2)\over 9(1-z)^2} 
  + 4\chi_2(1)\, {1 - 6z + z^2\over 9(1-z)^2} 
  - \frac43\, \chi_2'(1) \nonumber\\*
&+& \eta(1)\, {5 - 19z - 9z^2 - 25z^3 + 16z^4 \over 18(1-z)^3}
  - \eta'(1) {1 + z + 2z^2 \over 3(1-z)} \nonumber\\*
&+& \rho_0^2\, \Big[ \delta^{(1/m)}_{A_1}- \delta^{(1/m)}_{\FDs} 
  - \frac43\, \chi_2(1) \Big] \,.
\end{eqnarray}
Note that the coefficients of $\rho_0^2$ are independent of $\chi_i$.
While not all parameters entering these formulae are known, it may be
possible to get information on $\chi_2'(1)$ and $\eta''(1)$ from precise
measurements of $R_2'(1)$ and $R_1''(1)$, respectively, or from lattice QCD 
calculations.

Independent of model calculations, experimental data on the $B\to
D^*\ell\bar\nu$ form factor ratios defined in Eq.~(\ref{R12def}) will constrain
some of the subleading Isgur-Wise functions entering Eqs.~(\ref{delta1m}) and
(\ref{deltacm}).  Measurements of $R_{1,2}$ can be used to constrain the
quantities $\eta(1)$, $\eta'(1)$, and $\chi_2(1)$ according to
\begin{eqnarray}\label{R1exp}
R_1(1) &=& 1 + {4\bar\alpha_s\over 3\pi} 
  + {\bar\alpha_s^2\over \pi^2}\, \beta_0\,
  \bigg[ \frac29 - {1+z\over 3(1-z)}\, \ln z \bigg]
  + {\bar\Lambda\over 2m_c}\, \Big[ 1 + z - 2z\,\eta(1) \Big] 
  + \ldots \,, \nonumber\\
R_1'(1) &=& - {4\bar\alpha_s\over 9\pi} 
  - {\bar\alpha_s^2\over \pi^2}\, \beta_0\, \bigg[ {2(1+z^2)\over 9(1-z)^2} - 
  {(1+z)(1-4z+z^2)\over 9(1-z)^3}\, \ln z \bigg] \nonumber\\
&&{} - {\bar\Lambda\over 2m_c}\, \bigg[ {1+z\over 2} - z\,\eta(1) + 
  2z\,\eta'(1) \bigg] + \ldots \,.
\end{eqnarray}
Yet again, one encounters badly behaved perturbation series, $0.110 + 0.055$
for $R_1(1)$ and $-0.037 - 0.025$ for $R_1'(1)$.  For $R_2$ we obtain
\begin{eqnarray}\label{R2exp}
R_2(1) &=& 1 - {2\bar\alpha_s\over 3\pi}\, 
  \bigg[ {2z\over 1-z} + {z(1+z)\over (1-z)^2}\, \ln z \bigg] 
  - {13\bar\alpha_s^2\over 18\pi^2}\, \beta_0\, \bigg[ {z\over 1-z} +
  {z(1+z)\over 2(1-z)^2}\, \ln z \bigg] \nonumber\\*
&&{} - {\bar\Lambda\over 2m_c}\, 
  \Big[ (1 + 3z)\,\eta(1) + 4(1-z)\, \chi_2(1) \Big] + \ldots \,,
  \nonumber\\
R_2'(1) &=& - {2\bar\alpha_s\over 3\pi}\, 
  \bigg[ {z(1+10z+z^2)\over 3(1-z)^3} + {2z^2(1+z)\over (1-z)^4}\, \ln z \bigg]
  \\*
&&{} + {\bar\alpha_s^2\over 18\pi^2}\, \beta_0\, \bigg[ 
  {z(1-38z+z^2)\over 2(1-z)^3} +
  {z(1+z)(1-11z+z^2)\over (1-z)^4}\, \ln z \bigg] \nonumber\\*
&&{} + {\bar\Lambda\over 2m_c}\, 
  \bigg[ (1 + 3z) \bigg( {\eta(1)\over2} - \eta'(1) \bigg) - 
  4(1-z) \Big( \chi_2'(1) + \chi_2(1)\, \rho_0^2 \Big) \bigg] + \ldots \,. 
  \nonumber
\end{eqnarray}
Note that the ${\cal O}(\alpha_s)$ corrections to $R_2(1)$ and $R_2'(1)$ are
very small, around $+0.0056$ and $-0.0011$, respectively, and the ${\cal
O}(\alpha_s^2\beta_0)$ corrections are even smaller, $+0.0021$ and $-0.0006$,
respectively.  This is not unexpected, since all the nonperturbative
corrections to $R_2$ involve the subleading Isgur-Wise functions but not
$\bar\Lambda$ by itself, so the perturbation series does not involve a leading
renormalon.

\section{Discussion and Conclusions}

To evaluate the results of the previous section, they must be expressed in
terms of short distance quark masses.  We use the upsilon
expansion~\cite{upsexp}, and express $m_c$ through $m_b - m_c = \overline{m}_B
- \overline{m}_D + \lambda_1/2\overline{m}_B - \lambda_1/2\overline{m}_D$,
where $\overline{m}_B = (m_B+3m_{B^*})/4 = 5.313\,$GeV and $\overline{m}_D =
(m_D+3m_{D^*})/4 = 1.973\,$GeV.  It is convenient to re-express $\bar\alpha_s$
in terms of $\alpha_s(m_b)$.  For the $1S$ $b$ quark mass we use $m_b^{1S} =
4.75 \pm 0.07\,$GeV, which follows from Ref.~\cite{LLMW} using the CLEO
measurement of $\langle E_\gamma \rangle = 2.346 \pm 0.034\,$GeV in $B\to
X_s\gamma$ corresponding to a photon energy cut $E_\gamma >
2\,$GeV~\cite{CLEObsg}.  The uncertainty in $m_b^{1S}$, and the errors related
to it which we quote below, will almost certainly be significantly reduced in
the near future.  In the upsilon expansion
\begin{equation}
\bar\Lambda = m_B - m_b^{1S} - 0.051\, \epsilon 
  - 0.091\, \epsilon^2_{\rm BLM} + \ldots \,,
\end{equation}
where we used $\alpha_s(m_b) = 0.22$, $\epsilon \equiv 1$ is the parameter of
the upsilon expansion, and $\epsilon^2_{\rm BLM}$ denotes the part of the
second order correction proportional to $\beta_0$.  Since we include the
$\alpha_s^2\beta_0$ terms in all results, the residual scale dependence is
very small.

The right-hand sides of Eqs.~(\ref{delta1m}) and (\ref{deltacm}) can only be
estimated at present using model predictions.  The subleading Isgur-Wise
functions have been computed including order $\alpha_s$ corrections in QCD sum
rules, yielding~\cite{LNN}
\begin{equation}
\chi_2(1) \simeq -0.04\,, \qquad \chi_3'(1) \simeq 0.02\,, \qquad 
\eta(1) \simeq 0.6\,, \qquad \eta'(1) \simeq 0\,.
\end{equation}
Using the upsilon expansion to eliminate the quark masses, and these values for
the numerical estimates, we obtain
\begin{eqnarray}\label{slopefinal}
\rho_{\FD}^2 - \rho_{\FDs}^2 &=& 0.203 + 0.053\, \epsilon 
  - 0.013\, \epsilon^2_{\rm BLM} + 0.075\, \eta(1) + 0.14\, \eta'(1) 
  \nonumber\\*
&&{} + 1.0\, \chi_2(1) - 3.0\, \chi_3'(1) - 0.018\, \lambda_1/\mbox{GeV}^2
  \simeq 0.19 \,, \\*
\rho^2_{A_1} - \rho_{\FDs}^2 &=& 0.081 + 0.024\, \epsilon 
  - 0.006\, \epsilon^2_{\rm BLM}
  + 0.131 \eta(1) + 0.25 \chi_2(1) - 0.007\, \lambda_1/\mbox{GeV}^2
\simeq 0.17 \,, \nonumber
\end{eqnarray}
We also used $\lambda_1 = -0.25\,\mbox{GeV}^2$, although the results are hardly
sensitive to the value of this parameter.  The behavior of these perturbation
series are clearly much better than those in Eq.~(\ref{slopenum}) in terms of
the quark pole masses.  The uncertainty due to a $\pm70\,$MeV change in
$m_b^{1S}$ is $\mp 0.020$ and $\mp 0.022$ in these estimates of $\rho_{\FD}^2 -
\rho_{\FDs}^2$ and $\rho^2_{A_1} - \rho_{\FDs}^2$, respectively.

Although these estimates are model dependent, they are less so than one might
at first think, since the first terms on the right-hand sides of
Eqs.~(\ref{delta1m}), which are model independent, contribute a large part of
the result.  These results mostly depend on the value of $\eta(1)$ and on the
smallness of the functions $\chi_{2,3}(w)$, which parameterize order
$\lqcd/m_Q$ corrections due to the chromomagnetic operator.  If $\chi_2(1)$ and
$\chi_3'(1)$ were order unity then these results could be dramatically
different.  However, $\chi_{2,3}(w)$ are expected to be small in most models
(recall that $\chi_3(1) = 0$ due to Luke's theorem~\cite{Luke}), and as we will
see below, $\eta(1)$ can be constrained from $R_{1,2}(1)$.

For the curvature differences we obtain
\begin{eqnarray}\label{curvfinal}
c_{\FD} - c_{\FDs} &=& 0.202 + 0.050\, \epsilon - 0.012\, \epsilon^2_{\rm BLM}
  - 0.087\, \eta(1) - 0.08\, \eta'(1) - 0.07\, \eta''(1) \nonumber\\*
&&{} - 0.12\, \chi_2(1) - 1.0\, \chi_2'(1) + 1.5\, \chi_3''(1) 
  - 0.011\, \lambda_1/\mbox{GeV}^2 \nonumber\\*
&&{} + \rho_0^2\, [\rho_{\FD}^2 - \rho_{\FDs}^2 - 1.0\,\chi_2(1) 
  + 3.0\,\chi_3'(1) ] \simeq 0.17 + 0.29\, \rho_0^2 \,,\nonumber\\*
c_{A_1} - c_{\FDs} &=& 0.141 + 0.042\, \epsilon - 0.007\, \epsilon^2_{\rm BLM}
  - 0.059\, \eta(1) - 0.13\, \eta'(1) \nonumber\\*
&&{} - 0.12\, \chi_2(1) - 0.25\, \chi_2'(1) 
  - 0.005\, \lambda_1/\mbox{GeV}^2 \nonumber\\*
&&{} + \rho_0^2\, [ \rho_{A_1}^2 - \rho_{\FDs}^2 - 0.25\,\chi_2(1) ]
  \simeq 0.14 + 0.18\, \rho_0^2 \,, 
\end{eqnarray}
where for the numerical estimates we also used $\chi_2'(1) \simeq
0.03$~\cite{LNN}, and $\chi_3''(1) = \eta''(1) = 0$.  The uncertainty due to a
$\pm70\,$MeV change in $m_b^{1S}$ is $\mp 0.021$ and $\mp 0.018$ in the 0.17
and 0.14 terms in these estimates of $c_{\FD} - c_{\FDs}$ and $c_{A_1} -
c_{\FDs}$, respectively.  While there is a sizable uncertainty again due to the
subleading Isgur-Wise functions, for their particular values predicted by QCD
sum rules, the final result is dominated by terms which are model independent.

For the form factor ratios $R_1$ and $R_2$ we obtain
\begin{eqnarray}\label{R12final}
R_1(1) &=& 1.243 + 0.079\, \epsilon - 0.016\, \epsilon^2_{\rm BLM} 
  - 0.112\, \eta(1) - 0.021\, \lambda_1/\mbox{GeV}^2
  \simeq 1.25\,, \nonumber\\
R_1'(1) &=& -0.122 - 0.021 \epsilon + 0.010 \epsilon^2_{\rm BLM}
  + 0.056 \eta(1) - 0.112 \eta'(1) + 0.011 \lambda_1/\mbox{GeV}^2
  \simeq -0.10\,, \nonumber\\
R_2(1) &=& 1 + 0.006\, \epsilon + 0.001\, \epsilon^2_{\rm BLM} 
  - 0.355\, \eta(1) - 0.53\, \chi_2(1) \simeq 0.81\,, \\
R_2'(1) &=& - 0.001\, \epsilon + 0.178\, \eta(1) - 0.355\, \eta'(1) 
  - 0.53\, \chi_2'(1) - 0.53\, \chi_2(1)\, \rho_0^2 
  \simeq 0.09 + 0.02\,\rho_0^2\,. \nonumber
\end{eqnarray}
The uncertainty due to a $\pm70\,$MeV change in $m_b^{1S}$ is $\mp 0.03$, $\pm
0.01$, $\pm 0.03$, and $\mp 0.01$ in these estimates of $R_1(1)$, $R_1'(1)$,
$R_2(1)$, and $R_2'(1)$, respectively.  Unfortunately the sensitivity to the
values of the subleading Isgur-Wise functions is not too large, since they
enter with small coefficients.  Still, useful constraints can be obtained,
e.g., $R_2(1)$ measures $\eta(1)$ assuming that $\chi_2(1)$ is small (or a
linear combination of them otherwise), $\eta'(1)$ can be measured using
\begin{equation}
R_1(1) + 2R_1'(1) = 1 + 0.037\,\epsilon + 0.004\,\epsilon^2_{\rm BLM}
  - 0.223\,\eta'(1) \,,
\end{equation}
and then $R_2'(1)$ can be used to constrain a linear combination of $\chi_2(1)$
and $\chi_2'(1)$.  The coefficient of $\eta'(1)$ changes by $\pm 0.032$ under a
$\pm70\,$MeV variation of $m_b^{1S}$, while the other terms are essentially
unaffected.  Since $R_{1,2}''(1)$ are expected to be even smaller than
$R_{1,2}'(1)$, it seems unlikely that they could give useful independent
information.  CLEO measured $R_1 = 1.18 \pm 0.30 \pm 0.12$ and $R_2 = 0.71 \pm
0.22 \pm 0.07$~\cite{CLEOR12}, assuming that they are independent of $w$ and
that $h_{A_1}(w)$ has a linear $w$-dependence.  These results agree well with
Eq.~(\ref{R12final}).

Comparing our results for heavy quark symmetry breaking in the slope parameters
in Eq.~(\ref{slopefinal}) to the data in Table~I, there are several points to
be made: 

(i) The result of the linear fits at both CLEO and BELLE indicate
$\rho_{\FDs}^2 - \rho_{\FD}^2 \sim 0.2$ (see Table~I). This is opposite to what
is expected based on the QCD sum rule predictions for the subleading Isgur-Wise
function in Eq.~(\ref{slopefinal}).  The simplest way to accommodate the
central value of the data is if $\chi_3'(1) \sim 0.15$, which is several times
larger than the QCD sum rule prediction.  It should be straightforward to
decide using lattice QCD whether this large value of $\chi_3'(1)$ occurs.

(ii) The result of the unitarity constrained quadratic fits at both CLEO and
BELLE indicate $\rho_{A_1}^2 - \rho_{\FD}^2 \sim 0.3$ (see Table~I), whereas
based on the QCD sum rule predictions for the subleading Isgur-Wise functions
one would expect this difference to be close to zero,
\begin{equation}
\rho_{A_1}^2 - \rho_{\FD}^2 = -0.14 + 0.06\,\eta(1) - 0.14\,\eta'(1)
  - 0.75\,\chi_2(1) + 3.0\,\chi_3'(1) + \ldots \simeq -0.02 \,.
\end{equation}
The value $\chi_3'(1) \sim 0.15$ suggested above also accommodates this
data well.

(iii) The value of $\chi_3'(1)$ is hard to constrain from data, since its
contribution to $R_{1,2}'(1)$ is suppressed by $\alpha_s/\pi$.  A large value
of $\chi_3'(1)$ could explain a large heavy quark symmetry breaking in the
slope parameters, so determining $\chi_3'(1)$ from the lattice is very
important.  Of course, it would be desirable to compute all subleading
Isgur-Wise functions from the lattice.

(iv) Since heavy quark symmetry breaking would be unacceptably large in the
comparison of $\rho_{A_1}^2$ obtained from the unitarity constrained quadratic
fit with $\rho_{\FD}^2$ obtained from the linear fit, one must use the same
fitting procedure to extract the slopes and compare them. Therefore, one must
be careful not to draw wrong conclusions when comparing the two sides of
Fig.~1.\footnote{As the present authors first did.}  We expect the data for
$B\to D^*\ell\bar\nu$ will eventually favor the unitarity constrained fit, at
which point the data for $B\to D\ell\bar\nu$ will have to be fit with the same
method lest one gives up the symmetry relations that follow from the heavy
quark expansion or invokes surprisingly large values for subleading Isgur-Wise
functions (making convergence of the expansion questionable). 

(v) Measurements of heavy quark symmetry breaking in the slope and curvature
parameters, $\rho_X^2$ and $c_X$, together with measurements of the $R_{1,2}$
form factor ratios will strongly constrain the order $\lqcd/m_Q$ corrections. 
Eqs.~(\ref{slopefinal}), (\ref{curvfinal}), and (\ref{R12final}) can be used to
test future lattice calculations, or model predictions, such as those from QCD
sum rules used in this paper for some numerical estimates.

(vi) Most importantly, a better knowledge of the slope parameters will help to
reduce the error of $|V_{cb}|$, since there is a very strong correlation, as
can be seen from Fig.~1.

In conclusion, we calculated heavy quark symmetry breaking in the slopes and
curvatures of the $B\to D^{(*)}\ell\bar\nu$ spectra at zero recoil, including
the order $\alpha_s^2\beta_0$ corrections.  A combined fit to the shapes of the
$B\to D^*\ell\bar\nu$ and $B\to D\ell\bar\nu$ spectra together with the form
factor ratios $R_1$ and $R_2$ may lead to a better knowledge of the values of
the subleading Isgur-Wise functions.  This in turn may reduce the error of
$|V_{cb}|$, since the correlation between the $\rho_{\FDt}^2$ and $|V_{cb}|\,
\FDt(1)$ is very large.  Once $\rho_{\FD}^2 - \rho_{\FDs}^2$ is better
understood, it will also be interesting to see how well the $c_{\FD}-c_{\FDs}$
constraint is satisfied by the data fitted using the unitarity constraints.  
When $\FDt(w)$ is computed in unquenched lattice QCD, agreement of the
predicted values of $\rho_{\FD}^2 - \rho_{\FDs}^2$ and $c_{\FD}- c_{\FDs}$ with
data would give additional confidence (beyond checking that $|V_{cb}|$
extracted from $B\to D$ and $B\to D^*$ agree) that the errors are well
understood.  It will be especially reassuring if such a value of $|V_{cb}|$
agrees with the inclusive determination at the few percent level.

\begin{acknowledgments}

In the published article there is an error in the first relation in 
Eq.~(\ref{deltacm}), involving the terms proportional to $\rho_0^2$, slightly
affecting the numerical result in Eq.~(\ref{curvfinal}).  We thank Matt Dorsten
for communicating his results to us prior to
publication~\cite{Dorsten:2003ru}.  Erratum to be published in Phys.\ Lett.\ B.

We thank Mike Luke and Mark Wise for useful discussions, and Tom Browder,
Hyunki Jang, Jiwoo Nam and Karl Ecklund for correspondence about the BELLE and
CLEO analyses.
Z.L.\ thanks the Aspen Center for Physics for hospitality while part of this
work was completed.
B.G.\ was supported in part by the Department of Energy under contract No.\
DOE-FG03-97ER40546.
Z.L.\ was supported in part by the Director, Office of Science, Office of High
Energy and Nuclear Physics, Division of High Energy Physics, of the U.S.\
Department of Energy under Contract DE-AC03-76SF00098. 

\end{acknowledgments}

\end{document}